\documentstyle[amssymb,preprint,aps]{revtex}
\begin{document}
\title{Missing Mass, Dark Energy and the Acceleration of the Universe. Is
Acceleration Here to Stay?}
\author{Sel\c{c}uk \c{S}. Bay\i n}
\address{Middle East Technical University\\
Department of Physics\\
Ankara TURKEY\\
bayin@metu.edu.tr}
\date{\today }
\maketitle

\begin{abstract}
\end{abstract}

Recent measurements of the temperature fluctuations in the cosmic microwave
background radiation indicate that we live in an open universe. Size of
these fluctuations also indicate that the universe is almost flat. In terms
Friedmann models this implies a mass density within 10\% of the critical
density. However, the dynamical mass measurements\ can only account for
around 30\% of this mass. Recently, a series of outstanding observations
revealed that the cosmos is accelerating. This motivated some astronomers to
explain the missing 70\% as some exotic dark energy called the quintessence
or as the cosmological constant. In this paper we present an alternative
explanation to these cosmological issues in terms of the Friedmann
Thermodynamics. This model has the capability of making definite predictions
in-line with the current observations of the universe. According to this
model, cosmos was expanding slower at the beginning. During the galaxy
formation era; $z_{c}\in \lbrack 0.54,0.91],$ due to a change in the global
equation of state, it accelerates for a brief period of time. We expect to
see this as a discontinuity in the Hubble diagram. Recent data about the
galaxies with redshifts $0.5<$ $z<0.9$ displays this discontinuity clearly.
We expect the deceleration to re-appear as more data with redshifts $%
z\gtrsim 1$ is gathered. These galaxies will be among the very first
galaxies formed in the universe, thus still showing the kinematics of the
pre-galaxy formation era. This point is now clearly evidenced in the recent
data by Riess et al. on Type Ia supernovae with redshifts $z>1.25$ (2004
astro-ph/0402512). In our model, galaxies with redshifts $0\lesssim
z\lesssim 0.5$ should reflect the kinematics of the universe after the
transition is completed. These galaxies are now receding from each other
faster. However, for $\ z$ values \ towards the upper end of this range we
still expect to see deceleration. This is in contrast with the predictions
of the dark energy models. \newpage

\section{ Introduction}

One of the first applications of the Einstein's theory of gravitation was
given by Einstein to cosmology. Which later developed into what is now known
as the standard model, and has been extremely successful in explaining the
overall features of the universe to times as early as $10^{-2}\sec [1-3]$.
One of the basic features of the standard model ( also called the Friedmann
models) is the large scale homogeneity and isotropy of the universe. This is
evidenced in the uniformity of the temperature distribution of the CMBR
(cosmic microwave background radiation). Which is verified to one part in $%
10^{5}$ with antennas separated by angles ranging from 10 arcsec to 180
degrees. Among its other successful predictions; we could name the age of
the universe, existence of CMBR, expansion of the universe, abundances of
light elements, and the existence of structure. When the standard model is
extrapolated to the first microseconds of the universe, problems about the
horizon and flatness appear[1-3]. To solve these problems inflationary
models have been proposed. To explain the details of the universe up to the
Planck time of the origin may indeed be a tall order, even for a successful
theory like the Einstein's theory of gravity. However, there are also
serious problems regarding the relatively recent eras of the universe [1-3].

\subsubsection{Missing Mass Problem}

Recombination starts when the universe was about $300,000$ years old. Prior
to this time, light created with the big bang was constantly being scattered
by the free electrons in a plasma of primordial hydrogen and helium atoms.
At about this time, universe has sufficiently cooled for the electrons and
protons to form atoms hence, scattering has stopped. For this reason photons
that we see today as the CMBR at $2.73$ $K,$ carries information about the
state of the universe when it was only $300,000$ years old. However, CMBR is
not completely uniform. The theory predicts that there should exist
temperature fluctuations at the order of $10^{-5}$ in order to seed
structure formation. Indeed, such fluctuations have been observed by various
groups [4-9]. Since curvature acts like a lens, from the size of these \
fluctuations one could obtain valuable information about the geometry of the
universe. Compared to the critically open (flat) universe, these
fluctuations should appear smaller for a closed (spherical), and larger for
an open (hyperbolic) universe. Most recent observations [4] indicate that
the size of these fluctuations has not changed much since they were formed.
Thus indicating that the geometry of the universe is very close to flat.
Einstein's theory relates the matter content of the universe to its
geometry. Observations about the geometry of the universe, considered with
the Einstein's field equations and the Hubble constant $(H_{0})$
measurements, imply that the present density of mass should be within $10\%$
of the critical density defined as [4-11] 
\begin{equation}
\rho _{c}=\frac{3H_{0}^{2}}{8\pi G}.
\end{equation}
This result should naturally be confirmed by the dynamical mass measurements
in the universe. This is a challenging task. Various methods indicate that
only around 4\% of the matter is ordinary matter i.e. baryonic. From the
orbital speeds of galaxies within a cluster we also know that there is
approximately 6 times as much dark matter as baryonic matter. Dark matter is
basically composed of particles like neutrinos and other weakly interacting
massive particles (WIMPS). These particles interact weakly with other matter
and do not cluster. However, their presence could be detected through their
gravitational effects [4,12-14]. In other words, the total amount of all
kinds of matter -both dark and ordinary- in the universe only accounts for
30\% of the critical density. The remaining 70\% is far too large to be
missed by the current methods of detection, and yet it is still unaccounted
for. This is the so called 'missing mass' problem.

\subsubsection{Accelerating Universe?}

In the standard models, Einstein's theory presents three alternatives for
the geometry of the universe which depends on the density of the universe.
For densities above the critical value (1), universe is closed and the
geometry is spherical (Gaussian). For densities equal to the critical
density, the geometry is flat (Euclidean), and for densities less than the
critical density, universe is open and the geometry is hyperbolic
(Lobachevskian). In all these cases universe starts with a big bang, and the
geometry of the universe depends on the strength of the explosion i.e. the
total amount of energy (matter) in the universe.

Another property of the standard models is that for all conventional
equation of states which satisfy the inequality $(\rho +3P)\geq 0$, due to
the mutual gravitational attraction of different parts of the universe,
expansion decelerates [15a]. The specific amount of the deceleration depends
on the geometry, as well as the equation of state. In order to see this
deceleration, extending the Hubble diagram (velocity vs. distance or
redshift plot of galaxies) to higher and higher redshifts has been another
challenge for observational astronomy, and another means for determining the
geometry of the universe.

For nearby galaxies with redshifts 0.01-0.05, this relation is linear and
gives the value of the Hubble constant $H_{0}$ roughly as 66km/sec/Mpc
[8-10]. For higher redshifts we expect the graph to deviate from linearity
thus showing the effects of deceleration. In fact, just when we were
beginning to think that we are seeing the effects of deceleration [15b], new
observations with exciting and equally shocking results came in [16-18a,b].
For galaxies with redshifts 0.1 - 1, this data indicates that the universe
is accelerating? This is particularly surprising since acceleration starts
roughly at a time when the galaxies began forming: This is an era where we
have confidence in our theories of matter and Einstein's field equations.

\subsubsection{Inflation and Quintessence in a Nutshell}

Currently, popular solutions offered to the problems of the early universe,
as well as the missing mass and the accelerating universe problems are
related. They are all based on the revival of the cosmological constant
term, in one form or another. This term was originally introduced by
Einstein into his field equations to salvage his static universe model.
After the discovery of the expansion of the universe, he decided that it is
no longer needed and wanted to abandon it. However, since than this term has
become a center of attention in cosmology and kept appearing in Eintein's
equations for various reasons. Einstein expressed his dislike for this term
by saying '' the biggest blunder of my life '' and '' I introduced it, now I
can not get rid of it ''. Einstein's dislike for this term could hardly be
dismissed as purely emotional. If you consider it as a part of the left hand
side of the Einstein's field equations, it destroys the purely geometric
nature of this side. Besides, it adds a new arbitrary parameter to be
determined by observations, thus reducing the predictive power of the theory.

The other alternative is to consider it on the right hand side as a part of
the energy-momentum tensor. Now, the Lorentz covariance of this term makes
it convenient to interpret it as the energy-momentum tensor of the quantum
vacuum. When the cosmological term is compared with a perfect fluid, it has
the equation of state given as 
\begin{eqnarray}
P &=&-\rho  \nonumber \\
&=&-\Lambda .
\end{eqnarray}
Where, $\Lambda $ is the cosmological constant. This equation of state has
the unusual property of adding a repulsive force into the dynamics of the
universe. It is this property that is used in the inflationary models to
inflate the scale factor exponentially [1-3]. Thus, solving the horizon and
the flatness problems.

It is again this feature that is used for the missing mass and the
acceleration of the universe problems. However, if $\Lambda $ is taken as a
constant, it runs into the problem of fine tuning. To avoid this, it is
taken as a function of time, which is now interpreted as the density of a
new form of matter called 'quintessence' or dark energy [12,19]. This new
form of matter has positive energy and yet \ have the unusual property of
responding to gravity by repulsion, thus causing the acceleration. However,
not only its physical nature at the classical and the quantum levels is not
clear, it also requires rather special initial conditions to work.

In this paper, we present an alternative explanation to these cosmological
issues in terms of the Friedmann Thermodynamics. This model has the
capability of making definite predictions about the geometry of the
universe, the missing mass problem, and the acceleration of the universe,
all in-line with the current observations [20-26]. For future observations,
we also predict where this model will start differing from the dark energy
or the quintessence models. Models with the cosmological constant are also
referred to as dark energy (or quintessence) with constant density.

\section{Thermodynamics and Geometry}

Left hand side of the Einstein's field equations is a purely geometric term
constructed entirely from the metric tensor and its derivatives. However,
the right hand side i.e. the energy-momentum distribution of the universe,
which should describe the source of this geometry also contains the metric
tensor. Thus indicating that matter and geometry are interrelated in an
intricate way. For a given geometry, using the metric tensor Einstein's
equations could be used to obtain the total energy-momentum distribution of
the source. However, many different sources could be associated with a given
energy-momentum distribution. This indeterminacy about the details of the
source, which should be related to the information content of a given
geometry, immediately reminds us the entropy concept. In standard
statistical mechanics entropy is defined as proportional to the log of the
number of microstates which leads to the same macrostate. However, due to
the non-extensive nature of the self gravitating systems, even if we could
find a way to count the internal states of a given geometry, we could not
expect the corresponding 'curvature entropy' to be proportional to the log
of this number. A potential candidate may be the Tsallis' definition of
entropy $[27]$.

Another approach to search the connection between the geometry and
thermodynamics could be through the use of the second law, which states that
the total entropy of the universe can not decrease. However, due to the fact
that it is the total entropy that the second law is talking about, looking
for a geometric quantity that is an ever increasing function of time and
identifying it as the curvature entropy is not a reliable method. Besides,
for non-extensive systems the total entropy will not be a simple sum of its
components $[27]$. Thus, making the contribution coming from the geometry
even harder to identify. Considering these difficulties, we have recently
concentrated on the thermodynamic side of this problem and argued that a
system with finite 'curvature entropy' should also be endowed with a finite
'curvature temperature'. Being homogeneous and isotropic, Friedmann
geometries are ideal for searching this connection, where the methods of
equilibrium thermodynamics could still be used [20-26].

Starting point of the Friedman thermodynamics was the definition of the
'curvature temperature' as 
\begin{equation}
T=\alpha _{0}\left| \frac{k_{0}}{R(t)}\right| .
\end{equation}
$\alpha _{0}$ is a dimensional constant to be determined later and $R(t)$ is
the scale factor. $k_{0}^{2}/R(t)^{2}$ is proportional to the curvature
scalar of the constant time slices of the Friedman geometry ( $k_{0}^{2}=0$
for critically open i.e. flat, $k_{0}^{2}=1$ for closed, and $k_{0}^{2}=-1$
for the open universes.). As expected from a temperature like property, (1)
is uniform throughout the system and also a three-scalar. With this new
information (equation) added to the Einstein's field equations, and for a
'local' (flat spacetime) equation of state taken as 
\begin{equation}
P=\alpha \rho ,
\end{equation}
we were able to extract a 'global' equation of state, which now incorporates
the effects of curvature (temperature) as 
\begin{eqnarray}
\rho _{open}(T,P) &=&-\frac{c_{0}^{2}}{8\pi }(3+\frac{1}{\alpha })T^{2}+%
\frac{P}{\alpha }, \\
\rho _{closed}(T,P) &=&\frac{c_{0}^{2}}{8\pi }(3+\frac{1}{\alpha })T^{2}+%
\frac{P}{\alpha },\text{ \ \ \ \ \ }(c_{0}^{2}=\frac{4\pi ^{2}k^{2}c^{2}}{G%
\text{%
%TCIMACRO{\UNICODE{0x127}}%
%BeginExpansion
h\hskip-.2em\llap{\protect\rule[1.1ex]{.325em}{.1ex}}\hskip.2em%
%EndExpansion
}^{2}})\text{ }  \nonumber
\end{eqnarray}
for the open and closed models, respectively. These expressions, once
identified as the Gibbs energy densities, could be used to derive all the
required thermodynamic properties of the system. Note that $\rho $ and $P$
in (5) are no longer the same with their local values given in (4). They
reduce to their local values only in the ideal case where the geometry is
'exactly' flat. [20,21].

One remarkable consequence of this model is that one could now determine the
geometry of the universe by thermodynamic arguments. When we compare the two
geometries, we see that $\rho _{open}$ is always less than $\rho _{closed}$
i.e. 
\begin{equation}
\rho _{open}(T,P)-\rho _{closed}(T,P)\ =-\frac{2c_{0}^{2}}{8\pi }(3+\frac{1}{%
\alpha })T^{2}\langle 0
\end{equation}
Thus, making $\rho _{open}$ \ the more stable phase [20,21]. It is
interesting that recent observations on the inhomogeneities of the universal
background radiation, considered together with the dynamical mass
measurements also indicate that the universe is open. Even though it is very
close to the critically open i.e. flat case [4-9]. For the universe to be
'exactly' flat today, density has to be tuned to the critical density (1)
with infinite precession. This will make the flatness problem in the early
phases of the universe even more acute. Dynamical mass measurements can only
account for 30\% of the critical value [4].

In search for a justification of our definition of the curvature
temperature, we have studied Casimir effect in closed Friedmann models. By
taking the effective temperature of the Casimir energy as the curvature
temperature, we have identified the dimensional constant $\alpha _{0}$ as $\ 
\frac{1}{2\pi }\frac{\text{%
%TCIMACRO{\UNICODE{0x127}}%
%BeginExpansion
h\hskip-.2em\llap{\protect\rule[1.1ex]{.325em}{.1ex}}\hskip.2em%
%EndExpansion
c}}{k}.$ Later, by using the concept of local thermodynamic equilibrium, we
have extended our definition of curvature temperature to the sufficiently
slowly varying but otherwise arbitrary spacetimes [22,23]. When this
definition was used for spherically symmetric stars, we have shown that in
the black hole limit, the curvature temperature at the surface of the star
reduces to the Hawking temperature, precisely.

\section{Changes in the Local Equation of State and the \ Friedmann
Thermodynamics}

A large class of phase changes in the local matter distribution, including
the transition from the radiation to the matter era could be described as 
\begin{equation}
P=\alpha _{1}\rho \text{ \ }\rightarrow \text{ }P=\alpha _{2}\rho \text{ .}
\end{equation}
Aside from a change in the amount of deceleration, these transitions do not
lead to anything interesting within the context of standard Friedmann
models. However, considered in the light of Friedmann thermodynamics, they
offer new insights into some of the basic issues of cosmology.

We now concentrate on the beginning of the galaxy formation era, where the $%
\alpha $ value of the universe is expected to decrease. This follows from
the fact that at the onset of the galaxy formation, some of the gas in the
universe will be immobilized. Thus, giving less pressure for the same mean
density. For $P=\alpha \rho $, and an open universe the Gibbs energy density
was given in (3). For the transition $\alpha _{1}\rightarrow \alpha _{2},$
the difference between them could be written as 
\begin{equation}
\rho _{open,\alpha _{2}}(T,P)-\rho _{open,\alpha _{1}}(T,P)=\frac{%
c_{0}^{2}(\alpha _{2}-\alpha _{1})}{8\pi \alpha _{1}\alpha _{2}}[T^{2}-\frac{%
8\pi }{c_{0}^{2}}P].
\end{equation}
The two surfaces intersect along the curve 
\begin{equation}
T_{c}^{2}=\frac{8\pi }{c_{0}^{2}}P_{c}\text{ \ .}
\end{equation}
We could use the curvature temperature at the onset of the galaxy formation
era as the critical temperature $T_{c},$ and obtain $P_{c}$ from the above
relation. In ordinary phase transitions critical temperature is usually
defined with respect to the constant atmospheric pressure. In our case, at
the critical point both phases are expected to coexist, thus it is natural
to expect $P_{c}$ to lie somewhere in between the pressures just before the
transition has started, and after it has completed. In this regard, due to a
reduction in the local pressure, we expect $T^{2}-\frac{8\pi }{c_{0}^{2}}P<0$
before the critical point is reached, and $T^{2}-\frac{8\pi }{c_{0}^{2}}P>0$
after the transition is completed . Considering that $\alpha _{2}-\alpha
_{1}<0$, we could conclude that $\rho _{open,\alpha _{1}}(T,P)$ , and $\rho
_{open,\alpha _{2}}(T,P)$ are the stable phases before and after the
critical temperature, respectively.

\section{Dark Energy or the Missing Mass}

Now let us now see what new insights that this model contribute to
cosmology. Enthalpy density corresponding to the local equation of state $%
P=\alpha \rho ,$ could be written as 
\begin{equation}
h(s,P)=\frac{8\pi }{4c_{0}^{2}}(3+\frac{1}{\alpha })^{-1}s^{2}+\frac{P}{%
\alpha },
\end{equation}
where $s$ is the entropy density. During the phase transition ($\alpha
_{1}\rightarrow \alpha _{2})$ change in the enthalpy density could be
written as 
\begin{eqnarray}
\Delta h(s,P) &=&\frac{8\pi }{2c_{0}^{2}}(3+\frac{1}{\alpha })^{-1}s\Delta s+%
\frac{1}{\alpha }\Delta P,\text{ and} \\
\Delta h(s,P) &=&T\Delta s+\frac{1}{\alpha }\Delta P.
\end{eqnarray}
At constant pressure $\Delta h(s,P)$ gives us the energy density needed for
this phase transition. In ordinary phase transitions this energy would be
absorbed from a heat bath at constant temperature. In our case, since the
universe is a closed system, it could only come from within the system.
Calling this energy density $q_{c},$ we obtain it as 
\begin{equation}
q_{c}=\frac{2c_{0}^{2}}{8\pi }T_{c}^{2}\frac{(\alpha _{1}-\alpha _{2})}{%
\alpha _{1}\alpha _{2}},\text{ \ \ \ where}
\end{equation}
\begin{equation}
T_{c}=\frac{1}{2\pi }\frac{\text{%
%TCIMACRO{\UNICODE{0x127}}%
%BeginExpansion
h\hskip-.2em\llap{\protect\rule[1.1ex]{.325em}{.1ex}}\hskip.2em%
%EndExpansion
c}}{k}\frac{1}{R_{c}}.
\end{equation}
$R_{c}$ is the scale factor of the universe at the time of the transition. $%
q_{c}$ is the energy spent (used) by the system (universe) to perform the
above phase transition, which is required by the entropy criteria. In the
energy budget of the universe, this energy would show up as missing with
respect to the critically open (flat) case. To find how this energy would be
observed today, we use the scaling property of $q_{c}$, to obtain 
\begin{equation}
q_{now}=\frac{2}{8\pi }\frac{c^{4}}{G}\frac{(\alpha _{1}-\alpha _{2})}{%
\alpha _{1}\alpha _{2}}\frac{1}{R_{now}^{2}}.
\end{equation}

During the matter era baryons constitute the main source of pressure, whose
equation of state could be taken as the ideal gas law; 
\begin{equation}
P=[\frac{kT}{3\mu Hc^{2}}]\rho .\text{\ \ \ }
\end{equation}
\ $\rho $ is the energy density of baryons, $\mu $ is the mean molecular
weight and $H$ is the atomic mass unit. During the galaxy formation period,
which is expected to be short compared to the age of the universe, change in
temperature due to the expansion of the universe could be ignored. In other
words, we could treat $\alpha $ as a sufficiently slow varying parameter and
use equation (16) to assign an average value for it. Starting with the
recombination era, appearance of the first galaxies spans a temperature
range from\ several thousands to tens of K $[1-3]$. Taking 700K as an
average temperature and the mean molecular weight as 1, we obtain 2/3 as a
useful mean value for $\alpha $. However, even though the baryonic matter is
the main source of pressure, it is not the main source of matter. In terms
of the total energy density let us calculate an effective value for $\alpha $%
. We write the total local pressure as 
\begin{equation}
P_{total}=\overline{\alpha }_{1}\rho _{b}+\overline{\alpha }_{2}\rho _{nb}.
\end{equation}
First term represents the baryonic component, while the second represents
the nonbaryonic contribution. We take 
\begin{equation}
\overline{\alpha }_{1}=\frac{2}{3},
\end{equation}
and for weakly interacting particles we consider $\overline{\alpha }_{2}$ as
a number very close to zero. Expressing (17) as 
\begin{equation}
P_{total}=\frac{[\overline{\alpha }_{1}\rho _{b}+\overline{\alpha }_{2}\rho
_{nb}]}{\rho _{b}+\rho _{nb}}\rho _{total},
\end{equation}
\begin{equation}
P_{total}=[\overline{\alpha }_{1}(\frac{\ \rho _{b}}{\rho _{b}+\rho _{nb}})+%
\overline{\alpha }_{2}(\frac{\ \rho _{nb}}{\rho _{b}+\rho _{nb}})]\rho
_{total},\text{ and}
\end{equation}
using the fact that $\overline{\alpha }_{2}$ is a very small number, we
could introduce an effective $\alpha _{eff}$ \ value for the universe as 
\begin{equation}
P_{total}=\alpha _{eff}\rho _{total}\text{ ,}
\end{equation}
\begin{equation}
\alpha _{eff}=\overline{\alpha }_{1}(\frac{\ \rho _{b}}{\rho _{b}+\rho _{nb}}%
)\text{, where }\rho _{total}=\rho _{b}+\rho _{nb}\text{ .}
\end{equation}
Current observations [3,4,12] indicate that the present density of matter is
roughly distributed as:

\ \ \ \ \ \ \ \ \ \ \ \ \ \ \ \ \ \ \ \ \ \ \ \ \ \ \ 
\begin{tabular}{ll}
Radiation \ (photons) \ \ \  & 0.005\% \\ 
Ordinary visible dark matter (baryons) \  & \ 0.5\% \\ 
Ordinary nonluminous dark matter (baryons) & \ 3.5\% \\ 
Dark matter (WIMPS) & \ \ 26\%
\end{tabular}

Percentages are given \ with respect to the \ $\rho _{critical}$ (1).
Considering that roughly 10\% of the baryonic matter condenses in the form
of luminous matter thus decoupling from the expansion of the universe, we
could take the ranges of $\alpha _{1}$,and $\alpha _{2}$ as [3] 
\begin{eqnarray}
\alpha _{1} &\in &(\frac{4}{100},\frac{5}{100})\frac{2}{3}\rightarrow \alpha
_{2}\in (\frac{3.5}{100},\frac{4.5}{100})\frac{2}{3}\text{ , or} \\
\alpha _{1} &\in &(0.02667,0.03333)\rightarrow \alpha _{2}\in
(0.02333,0.03000)\text{ }  \nonumber
\end{eqnarray}
These percentages are consistent with the recent results from the cosmic
background imager (CBI) observations [4], and leads to 
\begin{equation}
\frac{(\alpha _{1}-\alpha _{2})}{\alpha _{1}\alpha _{2}}\in (3.33,\text{ }%
5.36)\text{ .}
\end{equation}
To calculate the range of $q_{now}$ , we take the value of $R_{now}$ as the
radius of the observable universe which could be taken as 
\begin{equation}
R_{now}\simeq 2\text{ x 10}^{10}ly\text{ = }2\text{ x 10}^{28}cm\text{ .}
\end{equation}
This gives $q_{now}$ in the range 
\begin{equation}
q_{now}\in (8.89\text{ x 10}^{-30}\text{, 1.43 x 10}^{-29})gm/cc.
\end{equation}
Cosmic microwave background radiation data is sometimes used to claim that
the geometry of the universe is flat (critically open). However, all it
actually says is that the geometry is open but very close to flat [4,11].
The 'huge' difference between the two cases and their potential cosequences
is usually overlooked [28]. For the universe to be \ exactly flat, its
density must be tuned to the critical density (1) with infinite precession.
From the recent data of \ VSA and CBI we could conclude that the density of
the universe is only within 10\% of the critical value [4,5 also see 11].
Considering that the observed matter density of the universe only adds up to
30\% of the critical density, the rest is declared either as 'missing', or
as exotic dark energy (quintessence) [12,19,28]. Recent data indicates that
the Hubble constant could be taken in the range [8-10]. 
\[
H_{0}\in \left( 55,75\right) km/\sec /Mpc. 
\]
This gives the range of the critical density ($\frac{3H_{0}^{2}}{3\pi G}$)
as 
\begin{equation}
\rho _{critical}\in \lbrack 6.05\text{ x}10^{-30},\text{ }1.44\text{ x}%
10^{-29}]gm/cc,
\end{equation}

\ \ Taking the missing mass as the 70\% of the critical density we find 
\begin{equation}
\rho _{\text{missing}}\in (4.23\text{ x 10}^{-30},1.01\text{ x 10}%
^{-29})gm/cc.
\end{equation}
$q_{now}$ is now well in the range given in (28). Certainly this energy does
not disappear from the universe, but it is needed for the phase transition,
and it is used for it.

This phase transition takes place during the formation of structure for the
first time in the universe. These are the first clusters, galaxies, quasars,
and superstars etc. Modern galaxies appear only during the last 10-15
billion years. During this era equation of state changes roughly from $P=%
\frac{1}{3}\rho $ to $P=0$ in a relatively short time compared to the age of
the universe. Thus, even a very crude approach like taking the arithmetic
mean of the $\alpha $ values of these equation of states for the average $%
\alpha _{2}$ value i.e. Taking 
\[
\alpha _{1}=1/3\text{ \ }\rightarrow \text{ }\alpha _{2}=1/6 
\]
gives $\frac{(\alpha _{1}-\alpha _{2})}{\alpha _{1}\alpha _{2}}=3$ , which
already \ leads to numbers that are very close to what we have obtained
before (22-28).

\section{Where did the Missing Mass Go ?}

To see how $q_{now}$ is spent, we write the free energy density and its
change as 
\begin{eqnarray}
f(T,v) &=&-\frac{c_{0}^{2}}{8\pi }(3+v)T^{2}, \\
\Delta f(T,v) &=&-\frac{c_{0}^{2}}{8\pi }2T(3+v)\Delta T-\frac{T^{2}c_{0}^{2}%
}{8\pi }\Delta v, \\
\Delta f(T,v) &=&-s\Delta T-P\Delta v.
\end{eqnarray}
For constant temperature processes, $\Delta f(T,v)$ would usually give the
work done by the system on the environment through the action of a boundary.
Since we have a closed system, this work is done by those parts of the
system expanding under its own internal pressure: 
\begin{equation}
w_{c}=P_{c}\Delta v=\frac{c_{0}^{2}T_{c}^{2}}{8\pi }\frac{(\alpha
_{1}-\alpha _{2})}{\alpha _{1}\alpha _{2}},
\end{equation}
where $w_{c}$ denotes the work done by the system at the time of the
transition. Now, at the critical point, $q_{c}$ amount of energy is used
from the system, while $w_{c}$ amount of it is used to do work to increase
the specific volume. In a closed system, we expect these two terms to cancel
each other. However, as opposed to the usually studied systems in
thermodynamics, where the changes take place infinitesimally slowly, our
system is dynamic i.e. The universe does not stop and go through this phase
transition infinitesimally slowly. As a result, we should also take into
account the change in the kinetic energy of the expansion. Hence, the energy
balance should be written as 
\begin{equation}
-q_{c}+w_{c}+\Delta (K.E.)_{c}=0.
\end{equation}
This implies 
\begin{equation}
\Delta (K.E.)_{c}=\text{ }\frac{c^{4}}{8\pi GR_{c}^{2}}\frac{(\alpha
_{1}-\alpha _{2})}{\alpha _{1}\alpha _{2}}>0\text{ \ \ (in ergs/cc),}
\end{equation}
which is the amount of energy that has gone into increasing the kinetic
energy of the expansion. In this model, $q_{c}$ amount of energy (density)
has been used (or converted) within the system. Part of it has gone into
work to increase the specific volume, while the rest is used to increase the
kinetic energy of the expansion. Why should the universe go through all this
trouble? Basically, for the same reason that water starts boiling when the
critical temperature is reached i.e. to increase its entropy.

\section{Location of the Critical Point}

To find the location of the critical point we first remember that in our
model the scale factor $R(t)$ (not the geometry) could be determined exactly
by using the local equation of state $P=\alpha \rho $ as $[19,20]$ 
\[
R(t)=\sqrt{C_{1}}\left[ \frac{3}{4}\left( \alpha +1\right) t+C_{0}\right] ^{%
\frac{2}{3(\alpha +1)}},\text{ where} 
\]
\begin{equation}
C_{0}=\frac{1}{2H_{0}}-\frac{3}{4}\left( \alpha +1\right) t_{0}\text{ ,}
\end{equation}
\[
\sqrt{C_{1}}=R_{0}(2H_{0})^{\frac{2}{3\left( \alpha +1\right) }}\text{.} 
\]
Using these we could write the ratio of the present value of the scale
factor to its value at the time of the transition as, 
\begin{equation}
(\frac{R_{0}}{R_{c}})^{2}=\frac{1}{\left[ \frac{3}{2}\left( \alpha
_{2}+1\right) H_{0}\left( t_{c}-t_{0}\right) +1\right] ^{\frac{4}{3(\alpha
_{2}+1)}}}\text{ .}
\end{equation}
Here, $t_{0}$ and $t_{c}$ represent the ages of the universe now and at the
time of the transition, respectively. Since redshift at the critical point
is given as 
\begin{equation}
z_{c}=\frac{R_{0}}{R_{c}}-1\text{ .}
\end{equation}
we could write 
\begin{equation}
z_{c}=\frac{1}{\left[ \frac{3}{4}\left( \alpha _{2}+1\right) 2H_{0}\left(
t_{c}-t_{0}\right) +1\right] ^{\frac{2}{3(\alpha _{2}+1)}}}-1\text{ .}
\end{equation}
In the light of the recent observations we could take the age $t_{0}$ and
the Hubble constant $H_{0}$ as 
\begin{eqnarray}
t_{0} &=&15\text{ x }10^{9}yrs\text{ ,} \\
H_{0} &=&60km/\sec /Mpc\text{ .}
\end{eqnarray}
These are among the most probable values for these parameters [9,10]. We
have also determined the range of $\alpha _{2}$ in equation$(23)$as 
\begin{equation}
\alpha _{2}\in (0.0233,0.03)\text{ .}
\end{equation}
For the age at the critical point we consider that first galaxies began
forming around 
\begin{equation}
t=1-2\text{ x }10^{9}yrs\text{ .}
\end{equation}
Naturally this transformation takes some time to be completed. Age of the
globular clusters is given around $\sim 13$ x 10$^{9}yrs.$ Thus we think
that it is reasonable to take [1] 
\begin{equation}
t_{c}\in \left( 2-5\right) 10^{9}yrs
\end{equation}
as the effective time of the transition. With these numbers we now obtain
the critical point in the range 
\begin{equation}
z_{c}\in \left( 0.54,0.91\right) \text{ .}
\end{equation}
This is consistend with the value $z\approx 0.73$ given by Perlmutter et al.
as the location of the cross-over point between deceleration and
acceleration [18a]. In our model cosmos was expanding slower at the
beginning. When the galaxy formation started at $z_{c},$ due to a change in
the global equation of state, it accelerates for a brief period of time. We
expect to see this as a discontinuity in the Hubble diagram, which is
usually plotted as relative intensity vs. redshift [16-18a,b,29]. Recent
data indicates that galaxies with redshifts $0.5<$ $z<0.9$ just began to
display the change in the Hubble parameter as our model predicts[16-18a,b].
We have mentioned that the deceleration should reappear as more data with
redshifts $z\gtrsim 1$ is gathered [26]. It is interesting to see that the
recent data obtained by Riess et al. clearly demonstrates this point [29].
These galaxies will be among the very first galaxies formed in the universe,
thus still showing the kinematics of the pre-galaxy formation era. Galaxies
with redshifts $0\lesssim z\lesssim 0.5$ should reflect the kinematics of
the universe after the transition. These galaxies are receding from each
other faster now, however for $\ z$ values towards the upper end of this
range we still expect to see deceleration. This is in contrast with the
predictions of the dark energy models, where the acceleration is forever
once the quintessence overtakes ordinary matter.

\section{Change in the Hubble Constant at the Critical Point}

Let us finally estimate the fractional change in the Hubble parameter at $%
z_{c}$. In Friedmann thermodynamics, local mass density calculated for a
region sufficiently small so that the effects of curvature could be
neglected was given as [20-25] 
\begin{equation}
\frac{8\pi G}{3c^{2}}\rho =\frac{\stackrel{.}{R}^{2}}{R^{2}c^{2}}.
\end{equation}
In terms of \ the Hubble parameter $H$ this could be written as 
\begin{equation}
\frac{8\pi G}{3c^{2}}\rho =\frac{H^{2}}{c^{2}},
\end{equation}
and the fractional change in $H$ could now be obtained as 
\begin{equation}
\frac{\delta H}{H}=\frac{8\pi G}{6H^{2}}\delta \rho .
\end{equation}
$\delta \rho $ is the energy used to increase the Hubble parameter (i.e. for
acceleration). At the critical point this was obtained as (34) thus, 
\begin{equation}
\delta \rho =\frac{c^{2}}{8\pi GR_{c}^{2}}\frac{(\alpha _{1}-\alpha _{2})}{%
\alpha _{1}\alpha _{2}}\text{, }
\end{equation}
and we obtain 
\begin{equation}
(\frac{\delta H}{H})_{c}=\frac{c^{2}}{6R_{c}}\frac{1}{H_{c}^{2}R_{c}}\frac{%
(\alpha _{1}-\alpha _{2})}{\alpha _{1}\alpha _{2}}.
\end{equation}
Using 
\begin{equation}
R_{c}=\frac{R_{0}}{1+z_{c}},
\end{equation}
and taking $R_{0}$ as $2$ x $10^{28}cm$ this could be written as 
\begin{equation}
(\frac{\delta H}{H})_{c}=[7.5\text{ x }10^{-9}\left( 1+z_{c}\right) ]\frac{1%
}{H_{c}^{2}R_{c}}\text{ }\frac{(\alpha _{1}-\alpha _{2})}{\alpha _{1}\alpha
_{2}}.
\end{equation}
\ \ \ 

Before we go any further we now follow a rather crude approach an obtain
another result for the fractional change in $H$. We take the average local
kinetic energy density of the expansion as 
\begin{equation}
K.E.=\frac{1}{2}\rho \stackrel{.}{R}^{2},
\end{equation}
we write its change $\Delta (K.E.)$ as 
\begin{equation}
\Delta (K.E.)=\rho R^{2}H^{2}\frac{\Delta H}{H}\text{ .}
\end{equation}
We have assumed $\Delta R\ll \Delta \stackrel{.}{R}$ during the transition.
At the time of the transition (actually after it has been completed) this is
equal to $\left( 34\right) $ thus, we could write 
\begin{equation}
(\frac{\Delta H}{H})_{c}=4.83\text{ x 10}^{47}\frac{1}{\rho
_{c}R_{c}^{4}H_{c}^{2}}\frac{(\alpha _{1}-\alpha _{2})}{\alpha _{1}\alpha
_{2}}\text{ .}
\end{equation}
In \ this equation subscript $c$ indicates the value of that parameter at $%
z_{c}$. Using the scaling property of $\rho $ as 
\begin{equation}
\frac{\rho _{o}}{\rho _{c}}=\frac{R_{c}^{3}}{R_{o}^{3}},
\end{equation}
we could write (54) as 
\begin{equation}
(\frac{\Delta H}{H})_{c}=4.83\text{ x 10}^{47}\frac{1}{\rho
_{o}R_{c}R_{o}^{3}H_{c}^{2}}\frac{(\alpha _{1}-\alpha _{2})}{\alpha
_{1}\alpha _{2}}.
\end{equation}
Using the values 
\begin{eqnarray}
\rho _{o} &=&7.16\text{ x }10^{-30}gm/cc,\text{ and} \\
R_{o} &=&2\text{ x }10^{28}cm  \nonumber
\end{eqnarray}
(56) becomes 
\begin{equation}
(\frac{\Delta H}{H})_{c}=8.43\text{ x 10}^{-9}\frac{1}{R_{c}H_{c}^{2}}\frac{%
(\alpha _{1}-\alpha _{2})}{\alpha _{1}\alpha _{2}}\text{ .}
\end{equation}
We could also use the relation (50), and 
\begin{equation}
H_{c}=H_{o}(1+z)^{\frac{3(\alpha _{2}+1)}{2}}
\end{equation}
to write expression $(54)$ entirely in terms of the present day values of $R$
and $H$. Thus using the ranges 
\begin{eqnarray}
\alpha _{1} &\in &(0.027,0.033)\text{, }\alpha _{2}\in (0.023,0.030)\text{,}
\nonumber \\
z_{c} &\in &(0.4516,0.54057)\text{, \ and taking} \\
H_{o} &=&60km/\sec /Mpc\text{,}  \nonumber
\end{eqnarray}
we obtain 
\begin{equation}
(\frac{\Delta H}{H})_{c}\in (14,\text{ }26)\%\text{.}
\end{equation}
\ Observationally we expect $(\frac{\Delta H}{H})_{c}$ to be among the
difficult parameters to determine. Since it gives a broader range for $(%
\frac{\Delta H}{H})_{c}$, we have used $(60)$ for $\ z_{c},$ which is
obtained by taking, 
\[
t_{c}\in (5-6\text{x}10^{9})yrs
\]
in (38).

In this model, effect of this phase transition will show up as a
discontinuity in the slope of the Hubble diagram roughly given by the amount
in (61). Considering the range of $\ z_{c}$ given in $(60),$ this result is
comparable to what $(51)$ would give. Using $(44)$ one obtains $(\frac{%
\Delta H}{H})_{c}\approx (7.39-15)\%.$

\newpage

\section{Conclusions}

Predictions of our model is in contrast with the predictions of the
quintessence models, where the acceleration starts around the galaxy
formation era but continues forever at an ever increasing pace $\left[ 12,19%
\right] $. This is due to the unusual nature of quintessence or dark energy
which responds to gravity by repulsion. Quintessence models, where the
density of dark energy remains constant are the models with the cosmological
constant. Recently, Wang and Tegmark have claimed that the CMBR data
actually favors the quintessence models with constant density i.e. the
cosmological constant models [28].

Due to the Lorentz covariant nature of its equation of state, cosmological
constant is usually interpreted as the quantum vacuum energy density. In
this case, as the universe expands, the amount of \ 'vacuum' (volume) and
thus the vacuum energy increases, while its density remains constant. In the
mean time, ordinary matter continues to thin out, thus increasing the effect
of repulsive force and the acceleration. Galaxy formation era is around
where the vacuum energy is expected to overtake ordinary matter. However, it
is not clear why the quantum vacuum energy should be Lorentz covariant. A
proper derivation of the renormalized quantum vacuum energy in curved
background geometries as the Casimir effect, gives a different result $%
[21-23]$. Actually, It is even difficult to philosophize about what 'pure'
vacuum- classical or quantum- should be. As soon as one considers the
presence of matter and/or observers, nature of the quantum vacuum energy
changes. Calculating the renormalized energy of the massless conformal
scalar field with a thermal spectrum at temperature T, in background closed
Friedmann geometry \ via the mode sum method, one sees that the quantum
vacuum energy gets completely washed out in the high temperature limit $(%
\frac{\text{%
%TCIMACRO{\UNICODE{0x127}}%
%BeginExpansion
h\hskip-.2em\llap{\protect\rule[1.1ex]{.325em}{.1ex}}\hskip.2em%
%EndExpansion
c}}{kRT}\ll 1)$, and is modified in the low temperature limit $(\frac{\text{%
%TCIMACRO{\UNICODE{0x127}}%
%BeginExpansion
h\hskip-.2em\llap{\protect\rule[1.1ex]{.325em}{.1ex}}\hskip.2em%
%EndExpansion
c}}{kRT}\gg 1)$. Considering that quintessence coexists with other matter
and the high temperature limit is the limit to be considered, interpretation
of quintessence as the quantum vacuum energy is bound to be problematic.
Indeed, a recent article by Ford discusses this point $[30].$

All three Friedman models are homogeneous and isotropic, and start with a
big bang. Critically open and the open universes are infinite in extent.
Hence, they start with an infinite amount of matter distributed uniformly
over an infinite space. However, due to the existence of particle horizon,
one could only observe a finite part of it. Thus, assertions about the
global topology of the universe are essentially very difficult to justify.
In this regard, all our arguments appearing in this paper are local and
independent of the global topology of the universe. In $[31],$ Gomero et al.
discusses problems regarding the observability of the global topology of the
universe. In our approach, we view geometry like different crystal
structures of matter i.e. matter distributed over different spaces
(lattices) with distinct symmetry properties. Thus, changes in symmetry are
allowed and interpreted as phase transitions $[20-26]$. However, It should
also be emphasized that in Friedmann thermodynamics topology does not have
to change. Indeed, for local equation of states given as $P=\alpha \rho $,
which covers a wide range of physically interesting cases, global topology
is always open (6). It would be interesting to see what \ kind of physically
acceptable local equation of states would induce such topology changes, if
at all possible.

Like the quintessence, Friedmann thermodynamics is also a suggestive model.
However, despite the missing pieces in its theoretical foundations, its
predictive power is incredibly high and not only it offers some very
interesting potential answers to the existing cosmological problems, but
also makes definite predictions for future observations. Other models
suggested for the accelerating universe and the dark energy problems could
be found in [32,33].

\newpage

\section{References}

[1]P. Coles, and F. Lucchin, Cosmology, John Wiley \& Sons Ltd. (2002).

[2]A.D. Dolgov, M.V. Sazhin, Ya.B. Zeldovich, Modern Cosmology, Editions
Frontiers(1990).

[3]E.W. Wolb, and M.S. Turner, The Early Universe, Addison-Wesley Publishing
Co.(1990).

[4]J.L. Sievers et al. arXiv:astro-ph/0205387 (2002).

[5]J. Silk, Physics World, {\bf 15}, no.8, pg.21 (2002).

[6]J. Silk, Physics World, {\bf 13}, no.6, pg.23 (2000).

[7]E. Cartlidge, Physics World, {\bf 14}, no.6, pg.5 (2001).

[8]R. Ellis, Physics World, {\bf 12}, no7, pg.19 (1999).

[9]G.A. Tammann et al., arXiv:astro-ph/0112489 (2001).

[10]W.L. Freedman, arXiv:astro-ph/0202006 (2002).

[11]\ M. Tegmark et al., arXiv:astro-ph/0310723 (2004).

[12]J. P. Ostriker, and P.J. Steinhardt, Sci. Am., Jan., pg.37 (2001).

[13]A. Taylor, and J. Peacock, Physics World, {\bf 14}, no.3, pg.37 (2001).

[14]N. Smith, and N. Spooner, Physics World, {\bf 13}, no.1, pg.23 (2000).

[15a]M.R. Robinson, The Cosmological Distance Ladder, W.H. Freemann and
Company NY (1985).

[15b]D. Goldsmith, Science, {\bf 276}, 37 (1997).

[16]C.J. Hogan, R.P. Kirschner, and N.B. Suntzeff, Sci. Am. Jan. pg.28
(!999).

[17]A.G. Riess, et al., Astronomical Journal, {\bf 116}, 1009 (1998).

[18a]S. Permutter, et al. arXiv:astro-ph/9812133 (1998).

[18b]R.A. Knop, Ap. J. {\bf 598}, 102 (2003).

[19]R. R. Caldwell, and P. J. Steinhardt, Physics World, {\bf 13}, no.11,
pg.31 (2000).

[20]S. Bayin, Ap. J.{\bf , 301}, 517 (1986).

[21]S. Bayin, Gen.Rel.Grav., {\bf 19}, 899{\bf \ (}1987{\bf ).}

[22]S. Bayin, Gen.Rel.Grav., {\bf 22},179{\bf \ (}1990{\bf ).}

[23]S. Bayin, Gen.Rel.Grav., {\bf 26},{\bf \ }951 (1994).

[24]S. Bayin, Proceedings of ECOS-95, pg.3, July11-15, Istanbul. Editors;

Y.A. Gogus, A. Ozturk, G.Tsatsaronis.

Also available at http://www.physics.metu.edu/\symbol{126}bayin .

[25]S. Bayin, Workshop on Second Law of Thermodynamics- Proceedings, Erciyes
Univ.

- T.I.B.T.D. 27-30/8/90 Kayseri(1990).

Also available at http://www.physics.metu.edu/\symbol{126}bayin .

[26]S. Bayin, \ IJMPD {\bf 11}, 1523: arXiv:astro-ph/0211097(2002).

[27]R. Slazar, and R. Toral, Physica, {\bf A290},159 (2001).

[28]Y.Wang and M. Tegmark, arXiv:astro-ph/0403292 (2004).

[29]A.G. Riess, et al., arXiv:astro-ph/0402512 (2004).

[30]L.H. Ford, arXiv:gr-qc/0210096v1(2002).

[31]G.I. Gomero, M.J. Rebu\c{c}as, R. Tvakal, arXiv:gr-qc/0210016v1 (2002).

[32]J. Ponce de Leon, arXiv:gr-qc/0401026v2 (2004).

[33]R.G. Vishwakarma and P. Singh, Class. Quan. Grav.20,2033 (2003):
arXiv:astro-ph/0211285v3 (2003).

\end{document}